\documentclass[prd,twocolumn,showpcs,amsmath,amssymb,nofootinbib,preprintnumbers,balancelastpage]{revtex4}
\pdfoutput=1
\usepackage{epsfig}
\usepackage{amsmath}
\usepackage{bm}
\usepackage{times}
\usepackage{graphicx}
\usepackage{color}
\usepackage{slashed}

\def\bea{\begin{eqnarray}}
\def\eea{\end{eqnarray}}
\def\ba{\begin{eqnarray}}
\def\ea{\end{eqnarray}}
\def\be{\begin{equation}}
\def\ee{\end{equation}}

\begin{document}
\preprint{Published in JHEP}

\title{Standard Model Vacua for Two-dimensional Compactifications}

\author{Jonathan M. Arnold, Bartosz Fornal and Mark B. Wise \\
\textit{California Institute of Technology, Pasadena, CA 91125, USA}\\
}
\date{\today}

\begin{abstract}
We examine the structure of lower-dimensional standard model vacua for two-dimensional compactifications (on a 2D torus and on a 2D sphere). In the case of the torus we find a new standard model vacuum
for a large range of neutrino masses consistent with experiment.
Quantum effects play a crucial role in the existence of this vacuum.
For the compactification on a sphere the classical terms dominate the effective potential for large radii and a stable vacuum is achieved only by introducing a large magnetic flux. We argue that there are no two-dimensional standard model vacua for compactifications on a surface of genus greater than one.
\end{abstract}

\maketitle
\bigskip

\section{Introduction}
While the standard model coupled to gravity has a unique four-dimensional vacuum, it has recently been shown \cite{Arkani} that it may also have a landscape of three-dimensional vacua. There might even exist vacua of lower-dimensionality. Transitions between such lower-dimensional vacua and the 4D standard model vacuum may have measurable effects on the cosmic microwave background radiation anisotropy (\cite{Graham,Salem,Adamek}; for further discussions on transitions between vacua of different dimensionality see \cite{Pillado,Sean}).

In this paper we focus on the vacua associated with two-dimensional compactifications of the standard model coupled to gravity (on a 2D torus and a 2D sphere). Their exact geometry is determined by competing contributions to the effective lower-dimensional potential. Classical contributions come from the four-dimensional cosmological constant and the curvature term resulting from dimensional reduction, whereas quantum contributions involve the Casimir energies of standard model particles. The basic formalism used in this paper was developed in \cite{Arkani} and this work extends \cite{Arkani} by explicitly finding new vacua for toroidal and spherical compactifications of the standard model. Our goal in this paper is to show by explicit computation that new standard model vacua arise with two compactified dimensions.

For the compactification on a 2D torus the only classical contribution to the effective potential is given by the 4D cosmological constant term from the 4D action. As such, the Casimir energies of standard model fields contribute significantly to the shape of the effective potential. We find that competitions between these contributions produce a new $AdS_2\times T^2$ standard model vacuum for a large range of neutrino masses and for a particular geometry of the toroidal space. For a discussion of the role of Casimir energies in higher-dimensional toroidal compactifications see for example \cite{6D1,6D2,6D3,6D4,Greene,BCPP,Ponton}.

The effective potential for the compactification on a 2D sphere, on the other hand, contains an extra classical term unique to compactifications on curved spaces. We find stable $AdS_2\times S^2$ vacua when the magnetic flux through the 2D sphere is large enough.

The work of this paper is not speculative. We determine some vacua of the standard model with two compactified dimensions using only long distance physics. Understanding the vacuum structure of the standard model is worthwhile even though the ultimate utility of this knowledge remains uncertain.

The results of this paper and \cite{Arkani} show that there are vacua of the standard model where either one or two dimensions are compactified. This opens the question: why did we end up cosmologically in a vacuum with three large spatial dimensions? Since the toroidal and spherical compactifications, which we find in this paper, correspond to an effective theory with one large spatial dimension and a negative cosmological constant, it is interesting to consider the rate for a tunneling process that creates a two-dimensional anti de Sitter bubble. Furthermore, since the new vacua are $AdS_3\times S^1$, $AdS_2\times T^2$, $AdS_2\times S^2$, the AdS/CFT correspondence may provide an alternative description of the standard model \cite{Arkani}.

\section{Compactification on a 2D torus}
Here we demonstrate the existence of lower-dimensional vacua of the standard model coupled to gravity in the case of a 2D compactification on a torus. We work in $N$ spacetime dimensions and consider the following spacetime interval,
\bea
ds^2 = g_{\mu\nu}(x) d x^\mu d x^\nu + t_{i j}(x) d y^i d y^j\,,
\eea
where $g_{\mu\nu}$ is the metric (gravitational convention) on the noncompact spacetime with $\mu, \nu = 0, 1, ..., N-3$ and $t_{i j}$ is the metric on the compact space with $i, j = 1, 2$. The compact coordinates are $y^i \in [0, 2\pi)$.

We parametrize the 2D torus with the metric
\bea\label{torus}
t_{i j} = \frac{a^2}{\tau_2}\left(
                              \begin{array}{cc}
                                1 & \tau_1 \\
                                \tau_1 & |\tau|^2 \\
                              \end{array}
                            \right) ,
\eea
where $\tau = \tau_1+ i \tau_2$ and $a^2$ are the shape and volume moduli, respectively, all being functions of the noncompact coordinates $x^\mu$.
The $N$-dimensional Einstein-Hilbert action is
\bea\label{4D-action}
\!\!\!\lefteqn{S = \int d^{N-2} x \,d^2 y \sqrt{- g_{(N-2)}\, t\,\,} }\nonumber\\
& & \ \ \ \times\left[\frac{1}{2}M_{(N)}^{N-2}{R}_{(N)}-\left(\rho_{(N)}+\Lambda_{(N)}\right)\right],
\eea
where $g_{(N-2)}={\rm det} (g_{\mu \nu})$,  $t={\rm det} (t_{i j})$ and $M_{(N)}$, $R_{(N)}$, $\rho_{(N)}$, $\Lambda_{(N)}$ are the $N$-dimensional Planck mass,  Ricci scalar, Casimir energy density, and cosmological constant, respectively. We note that
\bea
\Lambda_{(N)} &=& \Lambda_{(N)}^{\rm obs} - \Lambda_{(N)}^{\rm q. corr.}\ ,
\eea
where the superscript ``{\rm obs}'' indicates the observable value and ``{q.corr.}'' indicates the quantum correction in flat space.

After performing a reduction of the action from $N$ to $(N-2)$ dimensions using the explicit form of the 2D torus metric, we arrive at
\bea
\label{2D-action}
\lefteqn{\!\!\!\!\!S = \int d^{N-2} x \sqrt{- g_{(N-2)}}\bigg[(2\pi a)^2\frac{1}{2}M_{(N)}^{N-2}\bigg({R}_{(N-2)} }\nonumber\\
& &\ \ \!\!\!\!\!\ -\frac{(\partial_\mu \tau_1)^2+(\partial_\mu \tau_2)^2}{2\tau_2^2}+2\frac{(\partial_\mu a)^2}{a^2}\bigg)-V(a, \tau)\bigg],
\eea
where $V(a, \tau)=(2\pi a)^2\left(\rho_{(N)}+\Lambda_{(N)}\right)$. In order to find stationary points we perform a conformal transformation of the metric to the Einstein frame.  The desired transformation is
\bea
g_{\mu\nu} \rightarrow \left[\frac{(2\pi a)^2 \, M_{(N)}^{N-2}}{M_{(N-2)}^{N-4}}\right]^{-\frac{2}{N-4}} g_{\mu\nu}^{(E)}\ .
\eea
The action now takes the form
\bea\label{action}
\lefteqn{\!\!\!\!\!\!\!\!S=\int d^{N-2} x \sqrt{- g_{(N-2)}^{(E)}}\Bigg[\frac{1}{2}M_{(N-2)}^{N-4}\bigg({R}_{(N-2)}\,- }\nonumber\\
& & \!\!\!\!\!\!\!\!\!\!\!\!\!\!\! \frac{2(N-2)}{N-4}\frac{\left(\partial_\mu a \right)^2}{a^2} -\frac{(\partial_\mu \tau_1)^2+(\partial_\mu \tau_2)^2}{2\tau_2^2}\bigg) -V_{\rm eff}(a, \tau)\Bigg],
\eea
where
\bea
V_{\rm eff}(a, \tau)= \left[\frac{(2\pi a)^2 \, M_{(N)}^{N-2}}{M_{(N-2)}^{N-4}}\right]^{-\left(\frac{N-2}{N-4}\right)}V(a, \tau)\,.
\eea
Note that the conformal transformation does not exist for $N=4$, and we have to work in $(4+\epsilon)$ dimensions. We also assume $\epsilon > 0$; otherwise, the stationary point of action (\ref{action}) wouldn't be stable with respect to small spacetime dependent fluctuations of the volume modulus $a^2$.
The conditions for the existence of a stable stationary point can be easily obtained by minimizing the effective potential and taking the limit $\epsilon \rightarrow 0$. Those conditions are
\begin{eqnarray}\label{potentialtorus}
\lefteqn{V(a, \tau)=0\ , \ \ \ \ \partial_{\tau_{1,2}} V(a, \tau)=0\ , }\nonumber\\
 & &\ \partial_a V(a, \tau)<0\ ,\ \ \ \ \partial^2_{\tau_{1,2}} V(a, \tau)>0\ .
\end{eqnarray}
Their derivation is given in \cite{Arkani}. The equations of motion for the action (\ref{2D-action}) yield
\bea\label{Ricci}
4\,\pi^2 a\, M^{N-2}_{(N)} R_{(N-2)} = \partial_a V(a, \tau)
\eea
at the stationary point. This means that minima of the effective potential correspond to (two-dimensional) anti de Sitter vacua.

In the next step  we compute quantum contributions to the effective potential  using dimensional regularization. The Casimir energy density for a scalar of mass $m$ in an $N$-dimensional spacetime with $N-2$ dimensions flat and 2 dimensions compactified on a torus, assuming periodic boundary conditions, is\footnote{We neglect the very small curvature of the $(N-2)$-dimensional spacetime.}
\bea\label{sum}
\lefteqn{\!\!\!\!\!\!\!\!\!\!\!\!\rho_{(N)}(a, \tau_1, \tau_2, m)= }\nonumber\\
& & \!\!\!\!\!\!\!\!\!\!\! \!\!\!\!\!\!\!\!\frac{1}{(2\pi a)^2}\frac{1}{2}\sum_{n_1, n_2 = -\infty}^\infty \int \frac{d^{N-3} k}{(2\pi)^{N-3}} \sqrt{k^2 + t^{i j}n_i n_j+m^2}\, ,
\eea
where $t^{ij}$ is the inverse of $t_{ij}$ given in equation (\ref{torus}).
Equation (\ref{sum}) yields
\bea\label{A2}
\lefteqn{\!\!\!\!\!\!\!\!\!\!\!\rho_{(N)}(a, \tau_1, \tau_2, m)= }\nonumber\\
& &  \!\!\!\!\!\!\!\!\!\!\! \!\!\!\!\!\!\! -\frac{1}{(2\pi a)^2}\frac{\Gamma\left(-\frac{N-2}{2}\right)}{4\sqrt{\pi}(4\pi)^{\frac{N-3}{2}}}
\sum_{n_1, n_2=-\infty}^\infty\left(t^{i j}n_i n_j+m^2\right)^{\frac{N-2}{2}}.
\eea
The sum in the above equation is the extended Chowla-Selberg zeta function and can be written as (see, appendix \ref{app})
\bea\label{A3}
\lefteqn{\!\!\!\!\!\!\!\!\!\!\sum_{n_1, n_2=-\infty}^\infty\left(t^{i j}n_i n_j+m^2\right)^{\frac{N-2}{2}} }\nonumber\\
& &\!\!\!\!\!\!\!\!\!= m^{N-2} + \frac{1}{a^{N-2}}\bigg[2\tilde{\tau}_2^{\frac{2-N}{2}}\zeta_{\rm EH}\left(\tfrac{2-N}{2}, \tilde{\tau}_2 \,a^2 m^2\right)\nonumber\\
& &\!\!\!\!\!\!\!\!\! \ \  +\,2\sqrt{\pi}\,\frac{\Gamma\left(\frac{1-N}{2}\right)}{\Gamma\left(\frac{2-N}{2}\right)}\,\tilde{\tau}_2^{\frac{N}{2}}\zeta_{\rm EH}\left(\tfrac{1-N}{2}, \tfrac{a^2 m^2}{\tilde{\tau}_2}\right)\nonumber\\
& & \!\!\!\!\!\!\!\!\!\ \   +\,\frac{8\,\pi^{\frac{2-N}{2}}\sqrt{\tilde{\tau}_2}}{\Gamma\left(\frac{2-N}{2}\right)}\sum_{n, p=1}^\infty p^{\frac{1-N}{2}} \left(n^2 +\tfrac{a^2 m^2}{\tilde{\tau}_2}\right)^{\frac{N-1}{4}}\nonumber\\
& & \!\!\!\!\!\!\!\!\!\!  \ \ \times \cos(2\pi\,\tilde{\tau}_1 \,n \,p) K_{\frac{N-1}{2}}\left(2\pi\,\tilde{\tau}_2 \,p \sqrt{n^2+\tfrac{a^2 m^2}{\tilde{\tau}_2}}\right)\bigg],
\eea
where $\tilde{\tau}_i=\tau_i/|\tau|^2$ and $\zeta_{\rm EH}(s, q)$ is the Epstein-Hurwitz zeta function expressed by
\bea\label{A4}
\lefteqn{\!\!\!\!\!\!\!\!\!\zeta_{\rm EH}(s, q) = -\tfrac{1}{2}q^{-s} + \frac{\sqrt{\pi}}{2}\frac{\Gamma(s-\frac{1}{2})}{\Gamma(s)}q^{-s+\frac{1}{2}} }\nonumber\\
& & \ \ \ \ \ \ \ \ \ \ \ \ \ \!\!\!\!\!\!+
\frac{2\pi^s}{\Gamma(s)}q^{\frac{1-2s}{4}}\sum_{n=1}^\infty n^{s-\frac{1}{2}}K_{s-\frac{1}{2}}(2\pi n \sqrt{q})\ .
\eea
Note that
\bea
\Lambda_{(N)}^{\rm q.corr.} = - \frac{1}{2}(4\pi)^{-\frac{N}{2}}\Gamma\left(-\tfrac{N}{2}\right) m^N\,,
\eea
so that as $N\rightarrow 4$ the potential $V(a,\tau)$ can be expressed in terms of the observed cosmological constant $\Lambda_{(4)}^{\rm obs}$ and a finite quantum correction $\rho_{(4)}^{\rm obs}$\,,

\bea\label{rho}
\lefteqn{\!\rho_{(4)}^{\rm obs}(a, \tau_1, \tau_2, m) }\nonumber\\
& & \!\!= -\frac{1}{(2\pi a)^4}\Bigg[\frac{2(a \,m)^{3/2}}{\tilde{\tau}_2^{1/4}}\sum_{p=1}^\infty \frac{1}{p^{3/2}}K_{3/2}\left(2\pi \,p\, a \,m\sqrt{\tilde{\tau}_2}\right)\nonumber\\
& & \ \ \ \ \!\!+
\,2\,\tilde{\tau}_2 (a \,m)^2\sum_{p=1}^\infty \frac{1}{p^{2}}K_2\left(\tfrac{2\pi \,p\, a \,m}{\sqrt{\tilde{\tau}_2}}\right)\nonumber\\
& &\ \ \ \ \!\!+ \, 4 \sqrt{\tilde{\tau}_2} \sum_{n,p = 1}^{\infty} \frac{1}{p^{3/2}} \left(n^2 +\tfrac{(a \,m)^2}{\tilde{\tau}_2}\right)^{3/4}\nonumber\\
& & \ \ \ \ \!\!\times\cos(2\pi \,\tilde{\tau}_1\, p\, n)\, K_{3/2}\left(2\pi\, p \,\tilde{\tau}_2\sqrt{n^2 +\tfrac{(a \,m)^2}{\tilde{\tau}_2}}\right)\Bigg].
\eea
After assuming $|\tau|=1$  equation (\ref{rho}) coincides with the corresponding formula given in \cite{BCPP}.
In the massless case (\ref{rho}) reduces to
\bea
\lefteqn{\!\!\!\!\!\!\!\!\!\!\!\!\rho_{(4)}^{\rm obs}(a, \tau_1, \tau_2, 0) = -\frac{1}{(2\pi a)^4}\Bigg[\frac{\zeta(3)}{2\pi\tilde{\tau}_2} +\frac{\pi^2\tilde{\tau}_2^2}{90}\,+ }\nonumber\\
& & \!\!\!\!\!\!\!\!\!\!\!\!\!\!\!\! \,4\sqrt{\tilde{\tau}_2}\sum_{n,p = 1}^\infty \left(\frac{n}{p}\right)^{3/2}\cos(2\pi \,n\, p \,\tilde{\tau}_1) \, K_{3/2}(2\pi\, n\, p \,\tilde{\tau}_2) \Bigg].
\eea
Explicitly, the potential $V_1(a, \tau, m)$ for a single scalar can be written as
\bea
V_1(a, \tau, m) = (2\pi a)^2\left[\rho_{(4)}^{\rm obs}(a, \tau_1, \tau_2, m)+  \Lambda^{\rm obs}_{(4)}\right].
\eea
For the standard model we simply multiply $\rho_{(4)}^{\rm obs}(a, \tau_1, \tau_2, m)$ by the appropriate factor reflecting the number of degrees of freedom in four dimensions and add a minus sign for fermions
(i.e., $2$ for the photon, $2$ for the graviton, $-4$ for a Dirac neutrino, $-2$ for a Majorana neutrino) \cite{Arkani, Ponton, CPW, Rubin}
and sum over degrees of freedom.

Note that for $m \gg 1/a$ we obtain from equation (\ref{rho}) the relation $\rho_{(4)}^{\rm obs} \sim  \exp(-2\,\pi\, a \,m)$ (up to a factor of $\tilde{\tau}_2^{\pm 1/2}$ in the exponent).
We restrict our attention to the lengthscale $a \gg 1/m_e $ so that the Casimir energies of the electron and all heavier standard model particles are negligible compared to the contributions of the photon, graviton, and the neutrinos.\footnote{It will become clear from figures 1 and 2 that our results hold for the full range of $a$ where the standard model is valid.}

\subsection{Dirac neutrinos}
For the standard model with Dirac neutrinos the potential is
\bea\label{10}
\lefteqn{\!\!\!\!\!\!\!\!\!\!\!\!V(a, \tau) = (2\pi a)^2\bigg[\,4\,\rho^{\rm obs}_{(4)}(a, \tau_1, \tau_2, 0) }\nonumber\\
 & & \ \ \ \ \ - \,4\sum_{i=1}^3\rho^{\rm obs}_{(4)}(a, \tau_1, \tau_2, m_{\nu_i}) +  \Lambda^{\rm obs}_{(4)}\,\bigg].
\eea
Although the neutrino masses have not been determined, we can use experimental mass splittings for the atmospheric and solar neutrinos to generate the spectrum given the lightest neutrino mass and a choice of hierarchy. This allows us to investigate the potential for various values of the lightest neutrino mass. Experimentally, $\Delta m^2_{\rm atm} = (2.43 \pm 0.13)\times 10^{-3} {\ \rm eV^2}$, $\Delta m^2_{\rm sol} = (7.59\pm 0.20)\times 10^{-5} {\ \rm eV^2}$ \cite{pdg}.
We adopt the observational value of $\Lambda_{(4)}^{\rm obs} \simeq 3.1 \times 10^{-47} \ \rm GeV^4$ \cite{pdg} and numerically solve the three equations in (\ref{potentialtorus})
for $a$ with the formula for $V(a, \tau)$ given by equation (\ref{10}) and require the three inequalities in (\ref{potentialtorus}) to hold.
We denote the solution by $a_0$.
We narrow our analysis to the fundamental region $\ \tau_1 \in (-1/2, 1/2]$, $\ |\tau| \ge 1\ $ and $\tau_2>0$ \cite{6D4,Wilczek}, as this is the moduli space
of physically distinct vacua.

It turns out that even before performing the numerical analysis we can precisely determine the values of $\tau$ for which the potential (\ref{10}) has its extrema. Note that the Casimir energy density (\ref{sum}) is invariant under $\textrm{SL}(2, \mathbb{Z})$ modular transformations. In particular, the two generators of the modular group are $T:\tau\rightarrow\tau+1$, corresponding to a change of indices $(n_1, n_2) \rightarrow (n_1, n_2-n_1)$, and
$S:\tau\rightarrow -1/\tau$, being equivalent to $(n_1, n_2) \rightarrow (-n_2, n_1)$. The same symmetries are exhibited by the potential (\ref{10}) since it is a linear combination of the Casimir energies of the particles.
It has been argued that the fixed points $\tau = i$ (for the transformation $S$) and $\tau = 1/2+i\sqrt{3}/2$ (for $T S$) correspond to extrema of the potential \cite{6D4,Ponton,Wilczek}.
A numerical analysis shows that only the second one corresponds to a minimum of the potential. Thus, all new two-dimensional standard model vacua are characterized by
$\ (\tau_1, \tau_2) = (1/2, \sqrt{3}/2)$.

\begin{itemize}
\item[(a)] For a normal hierarchy the neutrino masses are
$(m_{\nu_1}^2, m_{\nu_2}^2, m_{\nu_3}^2) = (m_{\nu_1}^2,
m_{\nu_1}^2+\Delta m^2_{\rm sol},
m_{\nu_1}^2+\Delta m^2_{\rm atm}+\Delta m^2_{\rm sol})$.
There are no two-dimensional vacua for $m_{\nu_1}\lesssim 4.42 \times 10^{-12}\ {\rm GeV}$. For all masses $m_{\nu_1} \gtrsim 4.42 \times 10^{-12}\ {\rm GeV}$ we find precisely one stable vacuum.
Table I shows the values of $a_0^{\rm n.h.}$ for several masses $m_{\nu_1}$. The radii are expressed in $\rm GeV^{-1}$
$\left(10^{10}\ \rm GeV^{-1} \simeq 2 \ \mu m\right)$.
The plot of the potential $V(a, \tau)$ for $\ \tau = 1/2+i\sqrt{3}/2\ $ and several masses from table I is given in figure 1 (a).
\item[(b)] For an inverted hierarchy the neutrino masses are
$(m_{\nu_1}^2, m_{\nu_2}^2, m_{\nu_3}^2) = (m_{\nu_3}^2 + \Delta m^2_{\rm atm} - \Delta m^2_{\rm sol},
m_{\nu_3}^2+\Delta m^2_{\rm atm},
m_{\nu_3}^2)$.
This time there are no two-dimensional vacua for $m_{\nu_3} \lesssim 1.12 \times 10^{-12}\ {\rm GeV}$. For all masses $m_{\nu_3} \gtrsim 1.12 \times 10^{-12}\ {\rm GeV}$ we again find precisely one stable vacuum.
Values of $a_0^{\rm i.h.}$ for a few masses $m_{\nu_3}$ can be found in table I.
The plot of $V(a)$ for some of the masses from table I is given in figure 1 (b).
\end{itemize}
\begin{table}[h]
\label{table1}
\begin{center}
\begin{tabular}[t]{|c|c|c|c|c|c|}
  \hline
  $m_{\rm light\ \nu} {\ \left[{\scriptstyle \rm GeV}\right]}$ & \ $a_0^{\rm n.h.} {\ \left[{\scriptstyle \rm GeV^{-1}}\right]}$\
  & \ $a_0^{\rm i.h.} {\ \left[{\scriptstyle \rm GeV^{-1}}\right]}$\
  & $\ \ \ \ \ \ (\tau_1,\tau_2) \ \ \ \ \ \  $   \\ \hline \hline
  $\ \ \ 1.12\times 10^{-12}\ \ \ $ & --  &$2.54\times 10^{10}$ & $(1/2, \sqrt{3}/2)$  \\ \hline
  $4.42\times 10^{-12}$ & \ \ $4.76\times 10^{10}$ \ \ &\ \ $1.55\times 10^{10}\ \ $ & $(1/2, \sqrt{3}/2)$ \\ \hline
  $1.0\times 10^{-11}$ & $2.23\times 10^{10}$  &$1.21\times 10^{10}$ &  $(1/2, \sqrt{3}/2)$ \\ \hline
  $2.0\times 10^{-11}$ & $1.29\times 10^{10}$  &$9.28\times 10^{9}$ &  $(1/2, \sqrt{3}/2)$ \\ \hline
  $3.0\times 10^{-11}$ & $9.46\times 10^{9}$  &$7.66\times 10^{9}$ &  $(1/2, \sqrt{3}/2)$ \\ \hline
  $1.0\times 10^{-10}$ & $3.39\times 10^{9}$ &$3.28\times 10^{9}$ &  $(1/2, \sqrt{3}/2)$ \\ \hline
  $1.0\times 10^{-9}$ & $3.52\times 10^{8}$ &$3.52\times 10^{8}$ &  $(1/2, \sqrt{3}/2)$ \\ \hline
\end{tabular}
\end{center}
\vspace{-2mm}
\caption{\footnotesize{Values of $a_0^{\rm n.h.}, a_0^{\rm i.h.}, \tau_1, \tau_2$ for two-dimensional standard model vacua for several lightest Dirac neutrino masses $m_{ \rm light \ \nu}$ ($m_{\nu_1}$ for normal hierarchy, $m_{\nu_3}$ for inverted hierarchy). Note that $10^{10} {\ \rm GeV^{-1}} \simeq 2\ \mu m$.}}
\end{table}
\begin{figure}[h]
\vspace{2mm}
\centerline{\scalebox{0.9}{\includegraphics{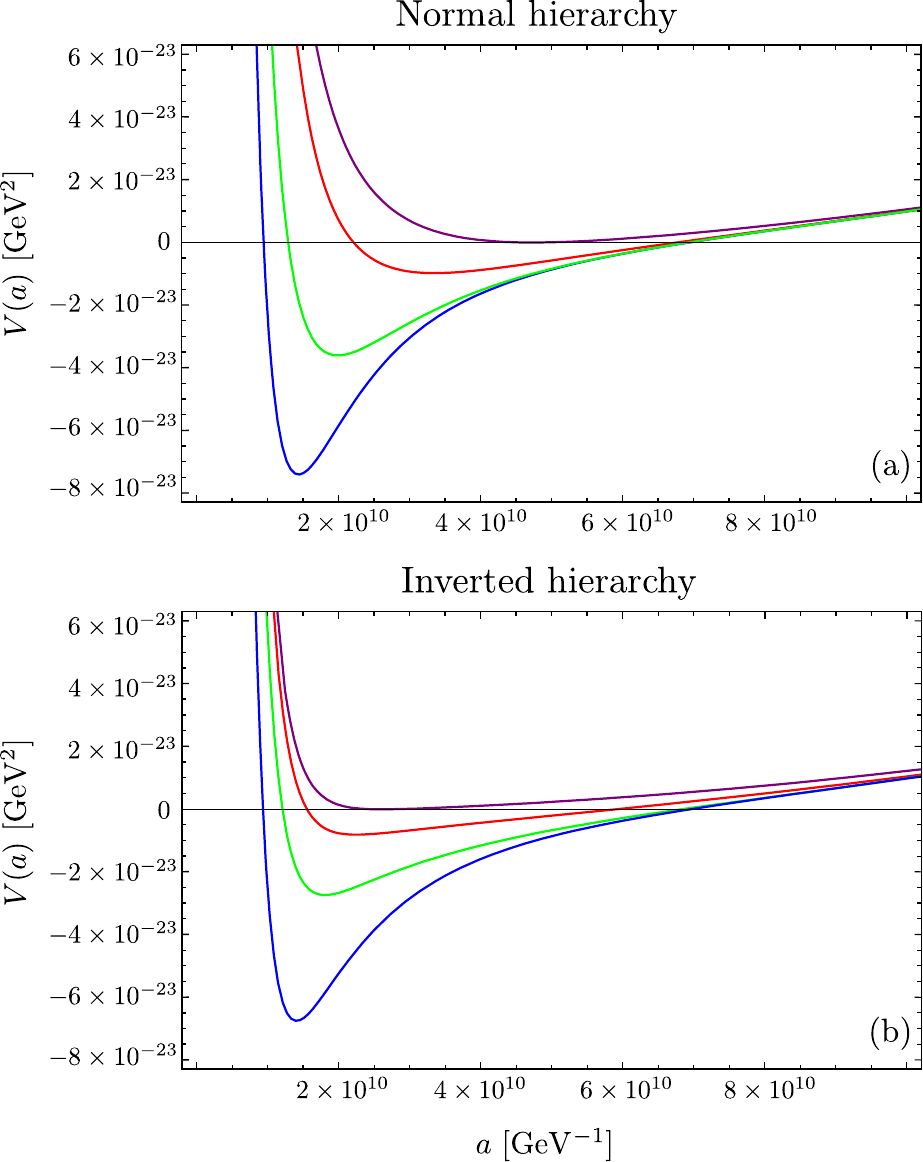}}}
\vspace{-2mm}
\caption{\footnotesize{(a) Plots of $V(a)$ for Dirac neutrinos with normal hierarchy for masses $m_{\nu_1} = 4.42 \times 10^{-12}\ {\rm GeV}$ (purple),
$1.0 \times 10^{-11}\ {\rm GeV}$ (red),  $2.0 \times 10^{-11}\ {\rm GeV}$ (green), and  $3.0 \times 10^{-11}\ {\rm GeV}$ (blue);
(b) Plots of $V(a)$ for Dirac neutrinos with inverted hierarchy for $m_{\nu_3} = 1.12 \times 10^{-12}\ {\rm GeV}$ (purple),
$4.42 \times 10^{-12}\ {\rm GeV}$ (red),  $1.0 \times 10^{-11}\ {\rm GeV}$ (green), and  $2.0 \times 10^{-11}\ {\rm GeV}$ (blue).}}
\end{figure}

\vspace{2cm}
\subsection{Majorana neutrinos}
In the case of the standard model with Majorana neutrinos the potential takes the form
\bea\label{11}
\lefteqn{\!\!\!\!\!\!\!\!\!\!\!\!V(a, \tau) = (2\pi a)^2 \bigg[\,4\,\rho^{\rm obs}_{(4)}(a, \tau_1, \tau_2, 0) }\nonumber\\
 & & \ \ \ \ \  - \,2\,\sum_{i=1}^3\rho^{\rm obs}_{(4)}(a, \tau_1, \tau_2, m_{\nu_i}) +  \Lambda^{\rm obs}_{(4)}\,\bigg].
\eea
In order to find new standard model vacua we perform analogous steps as in the case of Dirac neutrinos.
We find that for Majorana neutrinos there always exists precisely one standard model vacuum (this time for any value of $m_{\rm light \ \nu}$) characterized also by  $\ (\tau_1, \tau_2) = (1/2, \sqrt{3}/2)$.
Table II presents the values of $a_0$ for a few lightest neutrino masses for the normal and inverted hierarchies.
Figure 2 shows the plot of $V(a)$ for some of those masses.
\begin{table}[t]
\label{table1}
\begin{center}
\begin{tabular}[t]{|c|c|c|c|c|c|}
  \hline
  $\  m_{\rm light\ \nu} {\ [{\scriptstyle \rm GeV}]}\  $ & \ $a_0^{\rm n.h.} {\ \left[{\scriptstyle \rm GeV^{-1}}\right]}$\
  & \ $a_0^{\rm i.h.} {\ \left[{\scriptstyle \rm GeV^{-1}}\right]}$\
  &$\ \ \ \ \ \ (\tau_1,\tau_2) \ \ \ \ \ \  $  \\ \hline \hline
  $\ \ \ 0\ \ \ $ &$1.06\times 10^{10}$ &\ \ \ $5.17\times 10^{9}\ \ \ $ & $(1/2, \sqrt{3}/2)$  \\ \hline
  $\ \ \ 5.0\times 10^{-12}$\ \ \  & \ \ \ $9.29\times 10^{9}$ \ \ \ &\ \ $5.04\times 10^{9}\ \ $ & $(1/2, \sqrt{3}/2)$ \\ \hline
  $1.0\times 10^{-11}$ & $7.69\times 10^{9}$  &$4.79\times 10^{9}$ & $(1/2, \sqrt{3}/2)$  \\ \hline
  $2.0\times 10^{-11}$ & $5.64\times 10^{9}$  &$4.19\times 10^{9}$ &$(1/2, \sqrt{3}/2)$  \\ \hline
  $1.0\times 10^{-10}$ & $1.67\times 10^{9}$ &$1.62\times 10^{9}$ & $(1/2, \sqrt{3}/2)$  \\ \hline
  $1.0\times 10^{-9}$ & $1.74\times 10^{8}$ &$1.74\times 10^{8}$ & $(1/2, \sqrt{3}/2)$  \\ \hline
\end{tabular}
\end{center}
\vspace{-2mm}
\caption{\footnotesize{Values of $a_0^{\rm n.h.}, a_0^{\rm i.h.}, \tau_1, \tau_2$ for two-dimensional standard model vacua for several lightest Majorana neutrino masses $m_{\rm light\ \nu}$ ($m_{\nu_1}$ for normal hierarchy, $m_{\nu_3}$ for inverted hierarchy).}}
\end{table}
\begin{figure}[t]
\vspace{-4mm}
\centerline{\scalebox{0.9}{\includegraphics{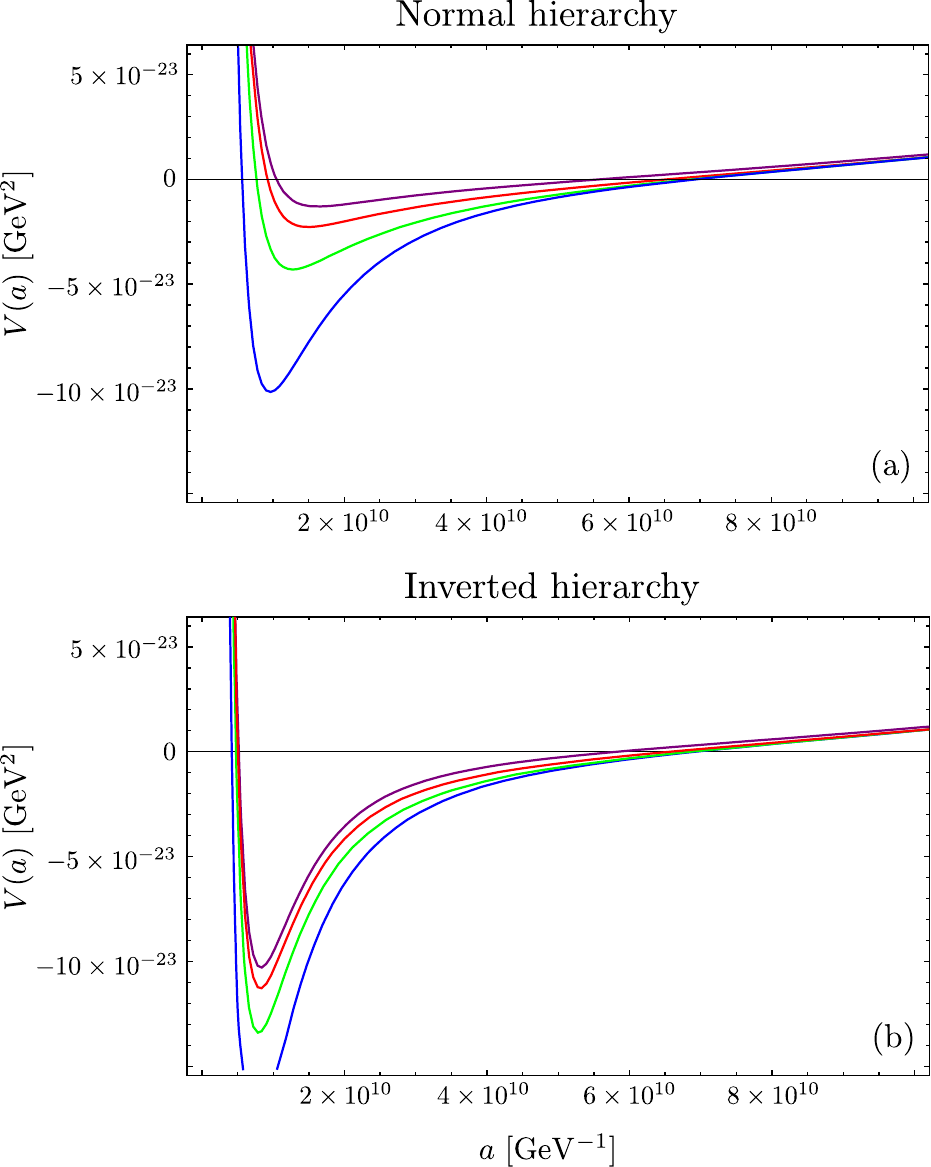}}}
\vspace{-2mm}
\caption{\footnotesize{(a) Plots of $V(a)$ for Majorana neutrinos with normal hierarchy for $m_{\nu_1} = 0$ (purple),
$5.0 \times 10^{-12}\ {\rm GeV}$ (red), $1.0\times  10^{-11}\ {\rm GeV}$ (green), and $2.0 \times 10^{-11}\ {\rm GeV}$ (blue);
(b) Plots of $V(a)$ for Majorana neutrinos with inverted hierarchy for $m_{\nu_3} = 0$ (purple),
$5.0 \times 10^{-12}\ {\rm GeV}$ (red), $1.0\times 10^{-11}\ {\rm GeV}$ (green),  and $2.0 \times 10^{-11}\ {\rm GeV}$ (blue).}}
\end{figure}

\section{Compactification on a 2D sphere}
Now we attempt to find the lower-dimensional standard model vacua for a 2D compactification on a sphere.
The spacetime interval we consider is
\bea
ds^2 = g_{\mu\nu}(x) d x^\mu d x^\nu + R^2(x) \left(d\theta^2 + \sin^2\theta \,d\phi^2\right)\,,
\eea
where $\theta\in[0, \pi],~ \phi \in [0, 2\pi)$.
The $N$-dimensional Einstein-Hilbert action is again given by equation (\ref{4D-action}) and, after the reduction to a $(N-2)$-dimensional spacetime, it takes the form
\bea
\label{2D-S-action}
\lefteqn{\!\!\!\!\!\!\!\!S = \int d^{N-2} x \sqrt{- g_{(N-2)}}\Bigg[(4\pi R^2)\frac{1}{2}M_{(N)}^{N-2} }\nonumber\\
& &\ \ \ \ \ \!\!\!\!\!\!\!\!\times\left({R}_{(N-2)}+2\, \frac{(\partial_\mu R)^2}{R^2}\right)-V(R)\Bigg],
\eea
where
\bea
V(R)= (4\pi R^2) \left[-\frac{M_{(N)}^{N-2}}{R^2}+\rho_{(N)}+\Lambda_{(N)}\right].
\eea
In the case of the sphere the potential includes not only the Casimir energy density and the cosmological constant term, but also the curvature of the sphere (in the case of the torus this additional term is obviously zero).

We again perform a conformal transformation of the metric and write down the action in the $(N-2)$-dimensional Einstein frame. The desired transformation is
\bea
g_{\mu\nu} \rightarrow \left[\frac{(4\pi R^2) \, M_{(N)}^{N-2}}{M_{(N-2)}^{N-4}}\right]^{-\frac{2}{N-4}} g_{\mu\nu}^{(E)}\ ,
\eea
and the action takes the form
\bea
\lefteqn{\!\!\!\!\!\!\!\!\!\!\!\!\!\!\!\!\!\!\!S =
\int d^{N-2} x \sqrt{- g_{(N-2)}^{(E)}}\Bigg[\frac{1}{2}M_{(N-2)}^{N-4}\bigg({R}_{(2+\epsilon)} }\nonumber\\
& &  \!\!\!\!\!\!\!\!\!\!\!\!\!\!\!\!\!\!\ \ \  - \frac{2(N-2)}{N-4}\frac{\left(\partial_\mu R \right)^2}{R^2}\bigg)-V_{\rm eff}(R)\Bigg],
\eea
where
\bea
V_{\rm eff}(R)= \left[\frac{(4\pi R^2) \, M_{(N)}^{N-2}}{M_{(N-2)}^{N-4}}\right]^{-\left(\frac{N-2}{N-4}\right)}V(R)\ .
\eea
The conditions for a stable stationary point are obtained similarly as in the torus case and, since this time the potential is a function only of $R$, they are
\bea
\label{potentialsphere}
V(R)=0\ , \ \ \ \ \partial_R V(R)<0\ .
\eea

Note that the 4-dimensional Planck mass is large $(M_{(4)} \simeq 1.22 \times 10^{19} {\ \rm GeV})$  and  the 4-dimensional cosmological constant is tiny $(\Lambda_{(4)}^{\rm obs} \simeq 3.1 \times 10^{-47} \ \rm GeV^4)$. We find that quantum corrections are negligible (with respect to both classical terms) for
$R \gg 1/(\Lambda_{(4)}^{\rm obs})^{1/4} \simeq 4\times 10^{11}{\ \rm GeV^{-1}}$. In the region of $R$ where classical effects dominate, the potential is
\bea\label{ps}
V(R)=4\pi\left[-M^2_{(4)}+R^2\Lambda_{(4)}^{\rm obs}\right].
\eea
The first condition in (\ref{potentialsphere}) yields the classical solution
\bea
R_0 = \frac{M_{(4)}}{\sqrt{\Lambda_{(4)}^{\rm obs}}} \simeq 4.3 \times 10^{26} {\ \rm m}\ ,
\eea
which is approximately the current Hubble radius, but this solution is unstable since the second condition is not fulfilled.

In the region $1/M_{(4)} \ll R \lesssim 1/(\Lambda_{(4)}^{\rm obs})^{1/4}$ the squared Planck mass still dominates the potential (the fact that quantum corrections become larger than the cosmological constant term is irrelevant) and there are no stable vacua since the equation $V(R)=0$ has no solutions.

However, it is possible to find stable vacua in the presence of a magnetic flux. This introduces an additional term in the potential (\ref{ps}) proportional to $ N^2/ (e R)^2$ (where $N$ is the number of magnetic flux quanta, $e$ is the electron charge) and yields exactly one stable solution
\bea
R = \left[\frac{M_{(4)}^2-\sqrt{M_{(4)}^4-\tfrac{N^2}{2e^2}\Lambda_{(4)}^{\rm obs}}}{2\Lambda_{(4)}^{\rm obs}}\right]^{1/2},
\eea
valid for $N<\sqrt{2}\,e \,M_{(4)}^2/(\Lambda_{(4)}^{\rm obs})^{1/2}$ and the region of $R$ where quantum corrections are negligible compared to both classical terms. The existence of a stable vacuum for $R \gg  4\times 10^{11}{\ \rm GeV^{-1}}$
requires $10^{30}\lesssim N \lesssim 10^{61}$, where the upper bound corresponds to $R \lesssim M_{(4)}/(2\Lambda_{(4)}^{\rm obs})^{1/2} \simeq 10^{42} {\ \rm GeV^{-1}}$.

Taking into account also the region of $R$ where just the squared Planck mass dominates over the quantum corrections, stable vacua
exist for the whole range $10^{17}  \lesssim N \lesssim 10^{61}$, where the lower bound
is set by the experimentally tested region of validity of the standard model, that is $R \gtrsim 1/(100 {\ \rm GeV})$.

We have also investigated 2D compactifications on surfaces of genus $g > 1$. The curvature of such manifolds is negative, therefore the squared Planck mass term enters the potential with a positive sign. In the region where quantum effects are negligible the equation for the volume modulus corresponding to the first condition in (\ref{potentialsphere}) has no solutions. This indicates that there are no 2D standard model vacua for such compactifications, even after introducing magnetic flux.

\section{Conclusions}
\vspace{-1mm}
We have investigated the vacuum structure of the standard model coupled to gravity with two spatial dimensions compactified (on a 2D torus and on a 2D sphere). We have calculated the potential, which, apart from the cosmological constant term, contains also Casimir energies of the standard model particles and, for the sphere case, includes a curvature term.

For the 2D torus compactification we did our analysis separately for the standard model with Dirac and Majorana neutrinos. For each of those cases we performed the calculation assuming a normal hierarchy, as well as an inverted hierarchy of the neutrino masses. We have shown that for the standard model with Majorana neutrinos there always exists precisely one new two-dimensional standard model vacuum. In the case of Dirac neutrinos we have also found exactly one standard model vacuum, but it exists only for a subset of the experimentally allowed neutrino masses.
The minimum value of the lightest Dirac neutrino mass for which there is a vacuum depends on the type of hierarchy.
The toroidal shape of the new vacuum for both Dirac and Majorana neutrinos is characterized by $(\tau_1, \tau_2) = (1/2, \sqrt{3}/2) $; its volume modulus $a^2$ depends on the mass of the lightest neutrino and we find $a$ to be on the order of microns. In our work we used periodic boundary conditions for the fermions. It is worthwhile to consider other boundary conditions.

The compactification on a 2D sphere is considerably different since the potential contains an additional curvature term resulting from dimensional reduction. For large enough radii the potential becomes classical. A straightforward classical analysis shows that a new stable two-dimensional standard model vacuum exists only after introducing a large magnetic flux. Finally, we have argued that for 2D compactifications on surfaces of genus $g > 1$ there are no new standard model vacua.

The calculations in this paper are not speculative physics. They contribute to determining the vacuum structure of the standard model and
depend only on long distance physics.

\subsection*{Acknowledgment}
\vspace{-2mm}
The work of the authors was supported in part by the U.S. Department of Energy under contract No. DE-FG02-92ER40701.

\appendix
\section{\textit{Generalized Chowla-Selberg formula}}
\label{app}
Here we derive, following the steps outlined in \cite{eli, elizalde3}, the formula for the regularized double sum in equation (\ref{A2}), in which the divergent pieces of the analytic continuation occur as poles in gamma functions for particular values of $N$. We start by writing down the formula for the Jacobi theta function
\bea\label{aa1}
\theta_3(z,w) = \sum_{n=-\infty}^{\infty}e^{i\pi(n^2 w +2 n z)}\,,
\eea
where $z$, $w$ are complex numbers and ${\rm Im}(w)>0$. We note that the following Jacobi identity holds \cite{abr}
\bea\label{aa2}
\theta_3\left(\tfrac{z}{w}, -\tfrac{1}{w}\right) = \sqrt{-i w}\,e^{i\pi \frac{z^2}{w}}\theta_3(z,w)\ .
\eea
After using (\ref{aa1}) to rewrite (\ref{aa2}) and expressing it in terms of the new variables \ $w'=-i \pi w$ \ and \ $z' = z/w$\  \ we arrive at
\bea\label{aa3}
\lefteqn{\sum_{n=-\infty}^{\infty}e^{-(n+z')^2 w'} }\nonumber\\
& & = \sqrt{\frac{\pi}{w'}}\left[1+2\sum_{n=1}^{\infty}e^{- \frac{\pi^2 n^2}{w'}}\cos(2\pi n z')\right]
\eea
with ${\rm Re}(w')>0$.  Notice that the corresponding formula in \cite{elizalde3} is missing a factor of 2.

We now express the general form of the double sum by
\bea
\lefteqn{\!\!\!\!\!\!\!\!\!\!\!\!\sum_{n, p = -\infty}^\infty\left(A p^2 + B p \,n + C n^2 +Q\right)^{-s} }\nonumber\\
& & \!\!\!\!\!\!\!\!\!\!\!\!\!\!\!\!\!\! = \frac{1}{\Gamma(s)}\sum_{n, p=-\infty}^\infty\int_0^\infty dt \,t^{s-1}e^{-(A p^2 + B p\, n + C n^2 +Q)t}\ .
\eea
We assume \ $A, \,Q > 0$\  and \ $4AC-B^2>0$.
Writing the quadratic form as
\bea
A p^2 + B p\, n + C n^2 = A\left(p+\tfrac{B}{2 A}n\right)^2+ \tfrac{\Delta}{4 A}n^2\ ,
\eea
where $\Delta = 4 A C - B^2$, and using relation (\ref{aa3}) with respect to the index $p$ for $z'=B \,n/(2A)$ and $w'= A \, t$, we arrive at
\bea\label{aa6}
\lefteqn{\!\!\!\!\!\!\!\!\!\!\!\!\sum_{n, p=-\infty}^\infty\left(A p^2 + B p\, n + C n^2 +Q\right)^{-s} }\nonumber\\
& & \!\!\!\!\!\! =\frac{1}{\Gamma(s)}\sum_{n=-\infty}^{\infty}\int_0^\infty dt\,t^{s-1}e^{-\left(\frac{\Delta}{4 A}n^2+Q\right)t}\nonumber\\
& & \times \sqrt{\tfrac{\pi}{A t}} \left[1+ 2\sum_{p=1}^\infty e^{-\frac{\pi^2 p^2}{ A t }}\cos\left(\pi \, n\, p \,\tfrac{B}{A}\right)\right].
\eea
The $n=0$ contribution to (\ref{aa6}) is
\bea
2A^{-s}\sum_{p=1}^\infty \left(p^2+\tfrac{Q}{A}\right)^{-s}+Q^{-s}\ .
\eea
Making use of the property of the modified Bessel functions of the second kind
\bea
\int_0^\infty du\,u^{s-1}e^{-\alpha^2 u - \frac{\beta^2}{ u}} = 2\left(\frac{\beta}{\alpha}\right)^s K_s(2\,\alpha\,\beta)\ ,
\eea
the contribution of the sum over non-zero $n$ gives
\bea
\lefteqn{\!\sum_{n, p=-\infty}^\infty\left(A p^2 + B p \,n + C n^2 +Q\right)^{-s}\Bigg|_{n\ne0} }\nonumber\\
& & \!\!\!\!\! =2\sqrt{\tfrac{\pi}{A}}\tfrac{\Gamma(s-\frac{1}{2})}{\Gamma(s)}\left(\tfrac{\Delta}{4 A}\right)^{\frac{1}{2}-s} \sum_{n=1}^\infty \left( n^2 +\tfrac{4A \,Q}{\Delta}\right)^{\frac{1}{2}-s}\nonumber\\
& &\!\!\!+\,\frac{8\pi^s}{\Gamma(s)}A^{-\frac{s}{2}-\frac{1}{4}}\left(\tfrac{\Delta}{4 A}\right)^{-\frac{s}{2}+\frac{1}{4}}\sum_{n, p=1}^{\infty} p^{s-\frac{1}{2}}
\left( n^2 +\tfrac{4A \,Q}{\Delta}\right)^{\frac{1}{4}-\frac{s}{2}}\nonumber\\
& &  \times\cos\left(\pi\,n\,p\,\tfrac{B}{A}\right)K_{s-\frac{1}{2}}\left(\tfrac{\pi\,p\sqrt{\Delta}}{A}\sqrt{n^2 + \tfrac{4A\,Q}{\Delta}}\right).
\eea
We point out that the corresponding formula in \cite{elizalde3} has an error in the definition of $b$ and is correct only after substituting $b\rightarrow 2b$.
We finally arrive at
\bea\label{app20}
\lefteqn{ \!\!\!\!\!\!\!\!\!\!\sum_{n, p=-\infty}^\infty\left(A p^2 + B p\, n + C n^2 +Q\right)^{-s} }\nonumber\\
& & \!\!\!\!\!\!\!\!\!\!= \,Q^{-s}+ 2A^{-s}\zeta_{\rm EH}\left(s, \tfrac{Q}{A}\right)\nonumber\\
& & \!\!\!\!\!+\, \frac{2^{2s}\sqrt{\pi}A^{s-1}}{\Delta^{s-\frac{1}{2}}}\frac{\Gamma(s-\frac{1}{2})}{\Gamma(s)}\zeta_{\rm EH}\left(s-\tfrac{1}{2}, \tfrac{4A\,Q}{\Delta}\right) \nonumber\\
& &\!\!\!\!\!+\,\frac{\pi^s}{\Gamma(s)}\frac{2^{s+\frac{5}{2}}\Delta^{\frac{1}{4}-\frac{s}{2}}}{\sqrt{A}}\sum_{n, p=1}^{\infty} p^{s-\frac{1}{2}}
\left( n^2 +\tfrac{4A \,Q}{\Delta}\right)^{\frac{1}{4}-\frac{s}{2}}\nonumber\\
& & \!\!\! \times\cos\left(\pi\,n\,p\,\tfrac{B}{A}\right)K_{s-\frac{1}{2}}\left(\tfrac{\pi\,p\sqrt{\Delta}}{A}\sqrt{n^2 + \tfrac{4A\,Q}{\Delta}}\right),
\eea
where the Epstein-Hurwitz zeta function
\bea
\zeta_{\rm EH}\left(s, q\right) = \sum_{n=1}^\infty\left(n^2+q\right)^{-s}
\eea
can also be regularized through the above procedure and
is given by formula (\ref{A4}).
From the form of the metric (\ref{torus}) and formula (\ref{A2}) we immediately identify
\bea\label{last}
\lefteqn{\!\!\! \!\!\!\!\!\!\!\!\!\! \!\!\!\!\!\!A = \frac{|\tau|^2}{a^2\tau_2}\ , \  \ B = - \frac{2\tau_1}{a^2\tau_2}\ , \  \ C=\frac{1}{a^2\tau_2}\ , }\nonumber\\
& & \!\!\!\!\!\! Q = m^2\ , \ \ s = \tfrac{2-N}{2}\ .
\eea
A quick check shows that $\Delta = 4/a^4 > 0$, thus formula (\ref{app20}) applies. Substitution of (\ref{last}) into (\ref{app20}) yields relation (\ref{A3}).


\end{document}